# Unidirectional magnetoresistance driven by nonequilibrium antiferromagnetic magnons

Xue He[1,‡], Hans Gløckner Giil[2,‡], Caiqiong Xu[1], Jicheng Wang[1], Arne Brataas[2], Jinbo Yang[3], Yanglong Hou[4,5], Rui Wu[1*], Shilei Ding[6†]

[1]*Spin-X Institute, School of Physics and Optoelectronics, State Key Laboratory of Luminescent Materials and Devices, and Guangdong-Hong Kong-Macao Joint Laboratory of Optoelectronic and Magnetic Functional Materials, South China University of Technology, Guangzhou 511442, China*

[2]*Center for Quantum Spintronics, Department of Physics, Norwegian University of Science and Technology, Trondheim 7491, Norway*

[3]*State Key Laboratory for Mesoscopic Physics, School of Physics, Peking University, Beijing 100871, China*

[4]*State Key Laboratory of Optoelectronic; School of Materials, Shenzhen Campus of Sun Yat-sen University, Shenzhen 518107, China*

[5]*School of Materials Science and Engineering, Peking University, Beijing 100871, China*

[6]*School of Physical and Mathematical Sciences, Nanyang Technological University, Singapore 637371, Singapore*

*Corresponding author: ruiwu001@scut.edu.cn;

†Corresponding author: shilei.ding@ntu.edu.sg.

‡These authors contributed equally to this work.

**Abstract**

Magnetoresistive effects are typically symmetric under magnetization reversal. However, nonlinear spin transport can give rise to unidirectional magnetoresistance in systems with strong spin-orbit interaction and broken inversion symmetry. Here, we demonstrate that the nonequilibrium magnon accumulation characterized by a finite magnon chemical potential can lead to a large and robust magnonic unidirectional spin Hall magnetoresistance (USMR) in the weakly coupled van der Waals antiferromagnet $CrPS_4$ in contact with Pt. Unlike conventional magnonic USMR driven by magnetization fluctuations, this effect persists under strong magnetic fields and low temperatures, with a pronounced peak near the spin-flip transition. The magnitude of magnonic USMR in $CrPS_4$/Pt exceeds that of YIG/Pt by more than two orders of magnitude and surpasses the electrical USMR in metallic Ta/Co bilayers by a factor of two. The observed field and temperature dependence indicates that spin transport is dominated by magnon chemical potential gradients rather than thermal- or fluctuation-driven magnon generation. These findings establish a new mechanism for nonlinear magnetoresistance in antiferromagnetic van der Waals heterostructures and open a route to magnon-based antiferromagnetic spintronic functionalities in two-terminal device geometries.

Magnetoresistance, which refers to the change in electrical resistance due to the orientation of magnetization, has long served as a foundational tool for probing spin-dependent transport in magnetic materials and also holds significant value in industrial applications due to its sensitivity and scalability [1,2]. Classic examples such as anisotropic magnetoresistance (AMR) [3], giant magnetoresistance (GMR) [4,5], and tunneling magnetoresistance (TMR) [6,7] are manifestations of spin-orbit coupling and spin-dependent conductivity in bulk ferromagnets, playing a pivotal role in both fundamental development of spintronics and technological advances. These effects typically exhibit symmetry under magnetization reversal and reflect a linear response with respect to the applied current. In recent years, nonlinear magnetoresistive effects [8-10] that break this symmetry have attracted growing attention for their potential to enable new functionalities in spintronic applications.

One significant example of nonlinear magnetoresistance is unidirectional spin Hall magnetoresistance (USMR), a two-terminal probe sensitive to the relative orientation between spin polarization and magnetization [11-13]. USMR arises from the asymmetric modulation of electrical resistance when a charge current generates a transverse spin accumulation via the spin Hall effect or Rashba-Edelstein effect, which subsequently interacts with an adjacent magnetic layer [14]. Being odd in both current and magnetization, USMR enables straightforward electrical readout of magnetic states and provides insight into spin-magnetization scattering and magnon-mediated interactions [15]. In metallic systems, USMR originates from multiple mechanisms, including spin-dependent scattering driven by interfacial spin accumulation [11,14-19] and field-induced distortions of the Fermi surface[20]. Additionally, magnon-mediated processes, termed as magnonic USMR, can influence resistance through spin current-induced magnon creation or annihilation, depending on the relative orientation between spin polarization and magnetization [15,21-29]. Especially, the magnonic USMR based on the magnetic insulators opens new possibilities for low joule heating and nonvolatile spintronic devices, but its practical implementation has so far been limited by weak signal amplitudes and strong suppression under external magnetic fields due to its sensitivity to spin current-induced magnetization fluctuations, as observed in systems such as YIG/Pt and $Fe_2O_3$/Pt [24,25,27]. This has motivated experimental efforts to explore alternative mechanisms for magnonic USMR, particularly those involving nonequilibrium magnon accumulation driven by spin current excited magnon populations, as indicated in the previous theoretical study [30], which may enable stronger and more robust magnonic USMR independent of magnetic fluctuations.

The weakly exchange coupled van der Waals antiferromagnet provides a compelling platform for realizing such mechanisms. Especially $CrPS_4$, whose crystal structure is composed of chromium octahedra linked via phosphorus atoms, forming magnetically coupled layers that exhibit A-type antiferromagnetic ordering below Néel temperature [31-33]. The weak interlayer exchange interaction results in low spin-flip field [34,35], and also leads to a low magnon gap, facilitating an efficient population of low-frequency GHz magnons [36]. These properties, along with its air stability and compatibility with nanoscale device fabrication [37], make $CrPS_4$ uniquely suited for investigating magnon-driven spin transport phenomena [38-41]. In particular, the low magnon gap improves sensitivity to spin current-induced magnonic phenomena, making $CrPS_4$ a promising candidate for realizing and modifying magnonic USMR.

In this work, we report a magnonic USMR originating from spin current-driven nonequilibrium magnon accumulation characterized by a finite magnon chemical potential, operating independently of magnetization fluctuations. The magnonic USMR observed in $CrPS_4$/Pt ($\Delta R_{USMR}/R_{\mathrm{xx}}^{1\omega}/j$ = $3.7 \pm 0.1 \times 10^{-16}$ $A^{-1}m^2$) exceeds that of YIG/Pt (~ $10^{-18}$) [24] by two orders of magnitude and is larger than previously reported electrical USMR in metallic ferromagnets such as Ta/Co (~ $10^{-16}$) [11]. Although its field dependence resembles that of the spin Seebeck effect, the underlying mechanism is fundamentally different. Rather than arising from thermally excited magnons generated by a temperature gradient, the signal originates from spatial gradients of the nonequilibrium magnon chemical potential induced by interfacial spin accumulation. As illustrated in Fig. 1, when an in-plane magnetic field is applied, the magnetic moments in the canted state tilt toward the external field. A nonequilibrium spin accumulation at the interface generates a magnon chemical potential gradient whose sign is set by the relative orientation between the injected spin polarization and the net magnetization. This gradient drives interfacial spin transport, producing spin Seebeck and magnon-mediated magnetoresistive effects.

Single-crystalline $CrPS_4$ was grown by chemical vapor transport using high-purity Cr, P, and S powders in a 1:1:4 ratio, with 5% excess sulfur as a transport agent. The mixture was sealed in a quartz ampoule and annealed in a two-zone furnace at 923 K (source) and 823 K (sink) for 7 days. The crystal structure was confirmed by X-ray diffraction and Raman spectroscopy [42,43], and magnetic properties were characterized by Physical Property Measurement System (PPMS, Quantum Design) equipped with a vibrating sample magnetometer (VSM) [42]. To prepare $CrPS_4$/Pt heterostructure, $CrPS_4$ flakes were mechanically exfoliated onto Si/$SiO_2$ substrates, followed by magnetron sputtering under a base pressure of $\sim 6 \times 10^{-8}$ Torr. Hall bar devices (10 μm × 13.7 μm) were fabricated via photolithography and ion beam etching. The thickness (~83 nm) of the flake was verified by atomic force microscopy (AFM) [42]. Electrical measurements were performed using PPMS. An alternating current of 0.5-1 mA at 13 Hz was supplied to the Hall bar or nonlocal device using a Keithley 6221 current source, and the resulting longitudinal and transverse voltage signals were detected with a lock-in amplifier (SR830).

Magnetic characterization of $CrPS_4$, shown in Fig. 2(a), confirms that it is a weakly exchange coupled antiferromagnet with spin-flop and spin-flip transitions observed at 0.8 T and 7.8 T, respectively, at 10 K. Temperature-dependent measurements on $CrPS_4$ flakes reveal a Néel temperature ($T_N$) of 38 K [42]. Fig. 2(b) shows a high-resolution transmission electron microscopy image of the $CrPS_4$ (104 nm)/Pt (22 nm) stack, revealing a clean and well-defined interface [42]. Such interfacial sharpness enhances spin transparency, facilitating efficient spin transport phenomena as evidenced by recent studies, including the spin Hall magnetoresistance [34], nonlocal spin transport [38-40], and spin Seebeck effect (SSE) [41].

We first demonstrate the SSE in a $CrPS_4$ (83 nm)/Pt (5 nm) bilayer. The optical micrograph of a $CrPS_4$/Pt Hall bar device and a schematic of the measurement configuration are shown in Fig. 2(c). An alternating current $\tilde{I} = I_0 \sin(\omega t)$ is applied through the Pt layer, inducing Joule heating and generating a vertical temperature gradient $\nabla T$ across the interface. This thermal gradient drives a spin current $\boldsymbol{J_s} = -S\nabla T$ in $CrPS_4$, where $S$ is the spin Seebeck coefficient. The thermally generated spin current is injected into the Pt layer, where it is converted into an electric field $\boldsymbol{E}_{\mathrm{ISHE}} \propto \theta_{\mathrm{SH}} \boldsymbol{J}_{\mathrm{s}} \times \boldsymbol{\sigma}$ via the inverse spin Hall effect [44]. Here, $\theta_{\mathrm{SH}}$ is the spin Hall angle of Pt, and $\boldsymbol{\sigma}$ is the spin polarization direction, which is aligned with the equilibrium magnetization $\boldsymbol{m}$ and/or Néel vector $\boldsymbol{n}$ [45]. The second-harmonic transverse resistance $R^{2\omega} = V^{2\omega}/I_0$ includes the thermally induced signal, as the temperature gradient scales with the square of the applied current $\tilde{I}^2 = {I_0}^2 \sin^2(\omega t)$. According to the symmetry of SSE, the SSE contribution to $R_{\mathrm{xy}}^{2\omega}$ (the SSE contribution to $R_{\mathrm{xx}}^{2\omega}$) reaches a maximum when the magnetic field is aligned along the *x*-axis (*y*-axis) and vanishes when the field is oriented along the *y*-axis (*x*-axis). Figs. 2(d) and 2(e) show the angular dependence of $R_{\mathrm{xy}}^{2\omega}$ and $R_{\mathrm{xx}}^{2\omega}$, respectively, measured in the *xy*-plane under a 9 T magnetic field at 5 K with an applied current of 1 mA. Such a high field can suppress the contributions from current-induced spin-orbit torque and Oersted fields to $R_{\mathrm{xy}}^{2\omega}$ and $R_{\mathrm{xx}}^{2\omega}$ [46]. Given the high resistivity of $CrPS_4$ at low temperatures [38], charge transport through the magnetic layer is negligible, effectively preventing thermal contributions from the Nernst effect. We therefore attribute the angular dependence of $R_{\mathrm{xy}}^{2\omega}$ in Fig. 2(d) to SSE [41], which is well described by a sine function $R_{\mathrm{xy}}^{2\omega} = \Delta R_{\mathrm{xy}}^{2\omega}\, sin\varphi$, where $\Delta R_{xy}^{2\omega} = 0.55 \pm 0.02$ mΩ is the magnitude of SSE measured from the transverse response, as shown by the red solid curve in Fig. 2(d).

The angular dependence of $R_{\mathrm{xx}}^{2\omega}$ in Fig. 2(e) also contains a contribution from SSE. Since the SSE voltage is proportional to the electrode separation along the direction of voltage detection (any anisotropy in SSE has been excluded using a device with symmetric cross geometry [42]), the magnitude of the SSE component in $R_{\mathrm{xx}}^{2\omega}$ can be expressed as $g\Delta R_{\mathrm{xy}}^{2\omega}$, where the geometric factor $g = 1.3$, depending on the geometry of the Hall bar, is further determined from the angular dependence of the first-harmonic resistance measurements [42,47]. The green dotted curve in Fig. 2(e) represents the SSE contribution, modeled as $g\Delta R_{\mathrm{xy}}^{2\omega}\cos\varphi$, which is clearly smaller than the measured angular dependence of $R_{\mathrm{xx}}^{2\omega}$. The remaining component is ascribed to the USMR, which follows the symmetry $R_{USMR} \propto \boldsymbol{J}_{\mathrm{c}} \cdot (\boldsymbol{m} \times \boldsymbol{z})$ [11]. The magnitude of USMR is determined to be $\Delta R_{USMR} = 0.22 \pm 0.02$ mΩ, with its angular dependence shown as the blue dash-dot curve in Fig. 2(e). Since $CrPS_4$ prohibits itinerant electron transport, conventional spin-dependent scattering mechanisms cannot account for the observed USMR. Instead, the effect is attributed to a magnonic origin, where spin current transportation at the interface plays a crucial role.

Figure 3(a) presents the magnetic field dependence of $\Delta R_{\mathrm{xx}}^{2\omega}$, $g\Delta R_{\mathrm{xy}}^{2\omega}$, and the extracted USMR component $\Delta R_{USMR} = \Delta R_{\mathrm{xx}}^{2\omega} - g\Delta R_{\mathrm{xy}}^{2\omega}$, obtained from angular-dependent measurements at various magnetic fields [42]. The SSE contribution to $\Delta R_{\mathrm{xx}}^{2\omega}$, represented by $g\Delta R_{\mathrm{xy}}^{2\omega}$, increases with the applied magnetic field and exhibits a peak near the spin-flip field, consistent with previous studies [41]. Given that the magnon gap in $CrPS_4$ is quite below 1 K in units of temperature, the field-induced modification of the magnon population via Zeeman splitting is negligible at higher temperatures. Therefore, the observed field dependence of the SSE is primarily governed by the canting of the magnetization, which increases with the field below the spin-flip threshold and reaches a maximum SSE signal at the spin-flip field [41].

A similar magnetic field dependence is observed in $\Delta R_{USMR}$ (as shown in Fig. 3(a)). This trend is further confirmed by direct field sweep measurements shown in Fig. 3(b), where the USMR signal reaches its maximum approaching the spin-flip field, as highlighted in Fig. 3(c). Previous studies have reported magnonic USMR in insulating systems, typically governed by spin current-induced magnetization fluctuations via spin-orbit torque [24,48]. In such scenarios, spin currents interact with local magnetization, resulting in magnon creation or annihilation depending on the relative orientation between spin polarization and magnetization [29]. These processes modulate the magnetization amplitude and thereby the spin transport at the interface, contributing to a nonlinear magnetoresistive signal [48]. However, the magnetization fluctuation-based mechanism is strongly reduced under applied magnetic fields as well as at low temperatures due to the suppression of thermal fluctuations [24]. This behavior contrasts with our observation of a pronounced magnonic USMR signal under such conditions, indicating that an alternative mechanism must be responsible. We note a recent study of USMR in $CrPS_4$/Pt shows a sign change across 4 T [49], which is not observed in our angular dependence and field dependence measurements [42].

Magnons, which are bosonic quasiparticles, do not have a conserved number, and their chemical potential is therefore zero. However, when magnon-magnon and magnon-phonon interactions occur on timescales much shorter than the magnon lifetime, the magnon system can be described by a quasi-equilibrium state [50]. In this regime, a nonequilibrium magnon chemical potential $\mu_\alpha$ and $\mu_\beta$ for magnon branch $\alpha$ and $\beta$ can be introduced to the Bose-Einstein distribution to describe the magnon population. In the $CrPS_4$/Pt heterostructure, spin current injection from the Pt layer via the spin Hall effect leads to magnon excitation in $CrPS_4$, establishing a spatially varying magnon chemical potential. This nonequilibrium magnon accumulation underlies key magnon-mediated spin transport phenomena, including non-local magnon transport [51,52] and spin Seebeck effect [53,54].

We calculate the interfacial spin current $J_s$ between Pt and $CrPS_4$ based on Fermi's golden rule for electron-magnon scattering processes at the antiferromagnet-normal metal interface [42]:

$$J_s = \left(J_m^{(0)} + J_m^{(1)} + J_m^{(2)}\right) sin\theta\, \boldsymbol{m} + \left(J_n^{(0)} + J_n^{(1)} + J_n^{(2)}\right) cos\theta\, \boldsymbol{n}, \quad (1)$$

$J_m^{(0)}(J_n^{(0)})$ corresponds to the spin Seebeck contribution from the temperature difference between the electron and magnon at the interface, $J_m^{(1)}(J_n^{(1)})$ is linear and $J_m^{(2)}(J_n^{(2)})$ is quadratic in terms of the $\mu_\alpha$ and $\mu_\beta$. The induced magnon chemical potential arises from the spin accumulation at the Pt interface and is therefore proportional to the magnitude of the applied charge current. $\theta$ is the canting angle of the $CrPS_4$ magnetization induced by the in-plane magnetic field. This angle is given by $\theta = arcsin\frac{\mu_0 H}{2\mu_0 H_E + \mu_0 H_A}$, where $\mu_0 H$, $\mu_0 H_E$, and $\mu_0 H_A$ represent the externally applied in-plane magnetic field, the interlayer exchange field, and the magnetic anisotropy field, respectively. Beyond the spin-flip transition, $\theta$ saturates at $\pi/2$.

Imposing continuity of the spin current at the interface requires $J_e^s(0) = J_s(0)$. Where $J_e^s(z)$ denotes the spatial distribution of the spin current in the Pt layer along the out-of-plane ($z$-axis) direction. The USMR is then defined through the resistivity difference under reversal of the applied electric field $E$:

$$\rho_{USMR} = (\rho(E) - \rho(-E))/2, \text{ where } \rho(E) = \frac{E}{\sigma E + \frac{2e}{\hbar}\theta_{SH} \int J_e^s(z) dz}, \quad (2)$$

Assuming the equal charge and magnon temperature, and thus vanishing $J_m^{(0)}(J_n^{(0)})$, the nonlinear (quadratic) terms in $J_e^s(z)$, ($J_m^{(2)}$ and $J_n^{(2)}$) contribute to a nonzero $\rho_{USMR}$. Considering $J^{(1)}_{m,n} \propto E$ and $J^{(2)}_{m,n} \propto E^2$, the overall resistivity exhibits a nonlinear dependence of the form $\rho(E) \propto \frac{E}{E+J^{(2)}_{m,n}} \sim \frac{E}{E+J^{(2)}_{m,n}} \sim J^{(2)}_{m,n}(\text{quadratic terms})/E \sim E,$ which gives rise to a finite USMR signal.

In the quasi-equilibrium regime, the nonequilibrium magnon population is described by a chemical potential shifted Bose-Einstein distribution. For small $\mu_{\alpha,\beta}$, the magnon density deviation is approximately $\frac{\partial n_B(\varepsilon)}{\partial \mu_{\alpha,\beta}}\mu_{\alpha,\beta}$, where $n_B(\varepsilon) = \frac{1}{e^{\frac{x}{k_BT}\pm\mu_{\alpha,\beta}}-1}$ is the Bose-Einstein distribution function. As the magnon gap is small, the suppression of magnon excitation due to magnetic field-induced gap opening remains weak in this field range, and the increase in $sin\theta$ dominates the field dependence of $J_s$. As a result, the interfacial spin current increases with the field below the spin-flip transition, thereby enhancing the USMR.

Above the spin-flip field, the magnetization becomes saturated and aligned with the field, such that $\theta = \pi/2$ remains constant. Further increases in the magnetic field no longer enhance the projection factor but raise the magnon gap in both $\alpha$ and $\beta$ modes. Consequently, the USMR exhibits a maximum near the spin-flip field and decreases at higher fields.

Figure 3(d) presents the normalized magnonic USMR signal, $\Delta R_{USMR}/R^{1\omega}_{\mathrm{xx}}$, as a function of applied current density, obtained from field-dependent measurements at 5 K [42]. The magnitude of the USMR is defined as $\Delta R_{USMR} = (R_{USMR}(8\,T) - R_{USMR}(-8\,T))/2$. The USMR exhibits a linear dependence on current density, consistent with the prediction from $J_s$. A linear fit yields the slope $\Delta R_{USMR}/R^{1\omega}_{\mathrm{xx}}/j$, which serves as a potential figure of merit for comparing USMR performance across different material systems. Notably, the extracted $\Delta R_{USMR}/R^{1\omega}_{\mathrm{xx}}/j$ in $CrPS_4$/Pt reaches $3.7 \pm 0.1 \times 10^{-16}$ $A^{-1}m^2$, which is at least twice as large as reported values in conventional Ta/Co bilayers ($\sim 10^{-16}$ $A^{-1}m^2$) [11]. Moreover, the observed magnonic USMR in $CrPS_4$/Pt exceeds that in YIG/Pt ($\sim 10^{-18}$ $A^{-1}m^2$) by more than two orders of magnitude [24].

To better understand the factors that affect the SSE and magnonic USMR, we performed temperature dependence measurements. When the applied magnetic field exceeds the spin-flip field, the sublattice magnetic moments align with the external field. Under this condition, the saturation magnetization $M_S$ can be extracted from magnetic hysteresis loop measurements [42].

Figures 4(a) and 4(b) show the temperature dependence of the $\Delta R_{\mathrm{xy}}^{2\omega}$ and $\Delta R_{USMR}/R_{xx}$, measured under an applied field of 8 T and a fixed current of 1 mA. Both signals vanish above the Néel temperature, consistent with the loss of long-range magnetic order. The disappearance of these signals reflects the suppression of magnon-mediated transport, since both SSE and magnonic USMR rely on thermally or spin current excited magnons. Upon decreasing the temperature below $T_N$, both quantities increase. However, the growth rate of the USMR, $\Delta R_{USMR}/R_{xx}$ is significantly larger than that of the SSE signal the $\Delta R_{\mathrm{xy}}^{2\omega}$. When the temperature is further reduced below approximately 5 K, the magnon are freezed out, leading to a suppression of both SSE and magnonic USMR.

The temperature dependence of the SSE is governed by the thermally excited magnon population described by the Bose-Einstein distribution and by the conversion of the resulting thermal spin current into a charge signal via the inverse spin Hall effect. A similar temperature dependence has been reported in $MnF_2$/Pt [55], where it is well explained by the combined effects of magnon population, magnon relaxation, and inverse spin Hall detection [56]. Both magnon population and relaxation play essential roles in determining the SSE magnitude, as they directly control the efficiency of magnon spin current injection into the Pt layer. While an increasing temperature enhances the thermal magnon population and thus the magnon spin current, it simultaneously increases magnon relaxation [57], which suppresses magnon transport. The competition between these two effects gives rise to a maximum in the SSE at an intermediate temperature.

Magnonic USMR is governed by intrinsically more complex spin-magnon coupling mechanisms. It relies on spin current injection into the magnetic layer to generate a nonequilibrium magnon chemical potential, as well as on the backflow of magnon spin current into the Pt layer, where the resistance is modified by additional charge currents generated through the inverse spin Hall effect. As a result, USMR is highly sensitive to magnon relaxation processes and exhibits a more pronounced increase upon decreasing temperature, as shown in Fig. 4(c).

In perspective, the magnonic USMR arises as a robust nonlinear spin transport effect driven by nonequilibrium magnon accumulation with a finite magnon chemical potential. Unlike fluctuation-based mechanisms that weaken under magnetic fields, the magnonic USMR in weak gap antiferromagnets grows with field and peaks near the spin-flip transition. The USMR observed in $CrPS_4$/Pt exceeds that of metallic ferromagnets and is orders of magnitude larger than fluctuation-driven signals in YIG/Pt. Its field and temperature dependence point to a mechanism dominated by spin current-induced nonequilibrium magnon transport. These findings position weakly exchange coupled van der Waals antiferromagnets as a powerful platform for exploring magnon chemical potential engineering and nonequilibrium magnon-mediated spin transport.

This work is supported by the National Key R&D Program of China (grant nos. 2022YFA1203902), the National Natural Science Foundation of China (NSFC) (grant nos. 12241401, 12374108), the Guangdong Provincial Quantum Science Strategic Initiative (grant no. GDZX2401002, GDZX2501003), the GJYC program of Guangzhou (grant no. 2024D01J0087), and the Research Council of Norway through its Center of Excellence 262633 “QuSpin”. S.D acknowledges support from the START-UP GRANT from Nanyang Technological University.

S.D. and R.W. conceived the experiments. X.H. synthesized the single crystal and fabricated the devices. X.H., S.D., C.X., J.W., and R.W. carried out transport and magnetic measurements. H.G.G. and A.B. contributed to the theoretical calculation. X.H., S.D., and R.W. contributed data analysis. S.D. drafted the manuscript, and all authors contributed to the reviewing and revising of the manuscript. Y.H., J.Y., and R.W. supervised the research, contributed to the acquisition of the financial support for the project leading to this publication.

The authors declare no competing interests.

## Reference

[1] J. Hu and T. F. Rosenbaum, Classical and quantum routes to linear magnetoresistance, Nat. Mater. **7**, 697 (2008).

[2] A. Fert, Nobel Lecture: Origin, development, and future of spintronics, Rev. Mod. Phys. **80**, 1517 (2008).

[3] T. McGuire and R. Potter, Anisotropic magnetoresistance in ferromagnetic 3d alloys, IEEE Trans. Magn. **11**, 1018 (1975).

[4] J. Q. Xiao, J. S. Jiang, and C. L. Chien, Giant magnetoresistance in nonmultilayer magnetic systems, Phys. Rev. Lett. **68**, 3749 (1992).

[5] M. N. Baibich, J. M. Broto, A. Fert, F. N. Van Dau, F. Petroff, P. Etienne, G. Creuzet, A. Friederich, and J. Chazelas, Giant Magnetoresistance of (001)Fe/(001)Cr Magnetic Superlattices, Phys. Rev. Lett. **61**, 2472 (1988).

[6] J. Mathon and A. Umerski, Theory of tunneling magnetoresistance of an epitaxial Fe/MgO/Fe(001) junction, Phys. Rev. B **63**, 220403 (2001).

[7] S. S. P. Parkin, C. Kaiser, A. Panchula, P. M. Rice, B. Hughes, M. Samant, and S.-H. Yang, Giant tunnelling magnetoresistance at room temperature with MgO (100) tunnel barriers, Nat. Mater. **3**, 862 (2004).

[8] T. Ideue, K. Hamamoto, S. Koshikawa, M. Ezawa, S. Shimizu, Y. Kaneko, Y. Tokura, N. Nagaosa, and Y. Iwasa, Bulk rectification effect in a polar semiconductor, Nat. Phys. **13**, 578 (2017).

[9] R. Wakatsuki and N. Nagaosa, Nonreciprocal Current in Noncentrosymmetric Rashba Superconductors, Phys. Rev. Lett. **121**, 026601 (2018).

[10] P. He, S. S. L. Zhang, D. Zhu, Y. Liu, Y. Wang, J. Yu, G. Vignale, and H. Yang, Bilinear magnetoelectric resistance as a probe of three-dimensional spin texture in topological surface states, Nat. Phys. **14**, 495 (2018).

[11] C. O. Avci, K. Garello, A. Ghosh, M. Gabureac, S. F. Alvarado, and P. Gambardella, Unidirectional spin Hall magnetoresistance in ferromagnet/normal metal bilayers, Nat. Phys. **11**, 570 (2015).

[12] R. Salikhov, I. Ilyakov, A. Reinold, J.-C. Deinert, T. V. A. G. d. Oliveira, A. Ponomaryov, G. L. Prajapati, P. Pilch, A. Ghalgaoui, M. Koch, J. Fassbender, J. Lindner, Z. Wang, and S. Kovalev, Ultrafast unidirectional spin Hall magnetoresistance driven by terahertz light field, Nat. Commun. **16**, 2249 (2025).

[13] I. H. Kao, J. Tang, G. C. Ortiz, M. Zhu, S. Yuan, R. Rao, J. Li, J. H. Edgar, J. Yan, D. G. Mandrus, K. Watanabe, T. Taniguchi, J. Hwang, R. Cheng, J. Katoch, and S. Singh, Unconventional unidirectional magnetoresistance in heterostructures of a topological semimetal and a ferromagnet, Nat. Mater. **24**, 1049 (2025).

[14] S. S. L. Zhang and G. Vignale, Theory of unidirectional spin Hall magnetoresistance in heavy-metal/ferromagnetic-metal bilayers, Phys. Rev. B **94**, 140411 (2016).

[15] C. O. Avci, J. Mendil, G. S. D. Beach, and P. Gambardella, Origins of the Unidirectional Spin Hall Magnetoresistance in Metallic Bilayers, Phys. Rev. Lett. **121**, 087207 (2018).

[16] C. O. Avci, K. Garello, J. Mendil, A. Ghosh, N. Blasakis, M. Gabureac, M. Trassin, M. Fiebig, and P. Gambardella, Magnetoresistance of heavy and light metal/ferromagnet bilayers, Appl. Phys. Lett. **107**, 192405 (2015).

[17] Y. Lv, J. Kally, D. Zhang, J. S. Lee, M. Jamali, N. Samarth, and J.-P. Wang, Unidirectional spin-Hall and Rashba−Edelstein magnetoresistance in topological insulator-ferromagnet layer heterostructures, Nat. Commun. **9**, 111 (2018).

[18] N. H. Duy Khang and P. N. Hai, Giant unidirectional spin Hall magnetoresistance in topological insulator – ferromagnetic semiconductor heterostructures, J. Appl. Phys. **126**, 233903 (2019).
[19] K. Hasegawa, T. Koyama, and D. Chiba, Enhanced unidirectional spin Hall magnetoresistance in a Pt/Co system with a Cu interlayer, Phys. Rev. B **103**, L020411 (2021).
[20] S. Shim, M. Mehraeen, J. Sklenar, J. Oh, J. Gibbons, H. Saglam, A. Hoffmann, S. S. L. Zhang, and N. Mason, Unidirectional Magnetoresistance in Antiferromagnet/Heavy-Metal Bilayers, Phys. Rev. X **12**, 021069 (2022).
[21] K. Yasuda, A. Tsukazaki, R. Yoshimi, K. S. Takahashi, M. Kawasaki, and Y. Tokura, Large Unidirectional Magnetoresistance in a Magnetic Topological Insulator, Phys. Rev. Lett. **117**, 127202 (2016).
[22] I. V. Borisenko, V. E. Demidov, S. Urazhdin, A. B. Rinkevich, and S. O. Demokritov, Relation between unidirectional spin Hall magnetoresistance and spin current-driven magnon generation, Appl. Phys. Lett. **113**, 62403 (2018).
[23] K.-J. Kim, T. Li, S. Kim, T. Moriyama, T. Koyama, D. Chiba, K.-J. Lee, H.-W. Lee, and T. Ono, Possible contribution of high-energy magnons to unidirectional magnetoresistance in metallic bilayers, Appl. Phys. Express **12**, 063001 (2019).
[24] G. Liu, X.-g. Wang, Z. Z. Luan, L. F. Zhou, S. Y. Xia, B. Yang, Y. Z. Tian, G.-h. Guo, J. Du, and D. Wu, Magnonic Unidirectional Spin Hall Magnetoresistance in a Heavy-Metal--Ferromagnetic-Insulator Bilayer, Phys. Rev. Lett. **127**, 207206 (2021).
[25] Y. Fan, P. Zhang, J. Han, Y. Lv, L. Liu, and J.-P. Wang, Observation of the Unidirectional Magnetoresistance in Antiferromagnetic Insulator $Fe_2O_3$/Pt Bilayers, Adv. Electron. Mater. **9**, 2300232 (2023).
[26] Z. Zheng, Y. Gu, Z. Zhang, X. Zhang, T. Zhao, H. Li, L. Ren, L. Jia, R. Xiao, H.-A. Zhou, Q. Zhang, S. Shi, Y. Zhang, C. Zhao, L. Shen, W. Zhao, and J. Chen, Coexistence of Magnon-Induced and Rashba-Induced Unidirectional Magnetoresistance in Antiferromagnets, Nano Lett. **23**, 6378 (2023).
[27] Y. Cheng, J. Tang, J. J. Michel, S. K. Chong, F. Yang, R. Cheng, and K. L. Wang, Unidirectional Spin Hall Magnetoresistance in Antiferromagnetic Heterostructures, Phys. Rev. Lett. **130**, 086703 (2023).
[28] V. E. Demidov, S. Urazhdin, E. R. J. Edwards, M. D. Stiles, R. D. McMichael, and S. O. Demokritov, Control of Magnetic Fluctuations by Spin Current, Phys. Rev. Lett. **107**, 107204 (2011).
[29] P. Noël, R. Schlitz, E. Karadža, C.-H. Lambert, L. Nessi, F. Binda, and P. Gambardella, Nonlinear Longitudinal and Transverse Magnetoresistances due to Current-Induced Magnon Creation-Annihilation Processes, Phys. Rev. Lett. **134**, 146701 (2025).
[30] W. P. Sterk, D. Peerlings, and R. A. Duine, Magnon contribution to unidirectional spin Hall magnetoresistance in ferromagnetic-insulator/heavy-metal bilayers, Phys. Rev. B **99**, 064438 (2019).
[31] S. Calder, A. V. Haglund, Y. Liu, D. M. Pajerowski, H. B. Cao, T. J. Williams, V. O. Garlea, and D. Mandrus, Magnetic structure and exchange interactions in the layered semiconductor $CrPS_4$, Phys. Rev. B **102**, 024408 (2020).
[32] Y. Peng, S. Ding, M. Cheng, Q. Hu, J. Yang, F. Wang, M. Xue, Z. Liu, Z. Lin, M. Avdeev, Y. Hou, W. Yang, Y. Zheng, and J. Yang, Magnetic Structure and Metamagnetic Transitions in the van der Waals Antiferromagnet $CrPS_4$, Adv. Mater. **32**, 2001200 (2020).
[33] Y.-X. Wang, T. K. M. Graham, R. Rama-Eiroa, M. A. Islam, M. H. Badarneh, R. Nunes Gontijo, G. P. Tiwari, T. Adhikari, X.-Y. Zhang, K. Watanabe, T. Taniguchi, C. Besson,

E. J. G. Santos, Z. Lin, and B. B. Zhou, Configurable antiferromagnetic domains and lateral exchange bias in atomically thin $CrPS_4$, Nat. Mater. **24**, 1414 (2025).
[34] R. Wu, A. Ross, S. Ding, Y. Peng, F. He, Y. Ren, R. Lebrun, Y. Wu, Z. Wang, J. Yang, A. Brataas, and M. Kläui, Magnetotransport Study of van der Waals $CrPS_4$/(Pt,Pd) Heterostructures: Spin-Flop Transition and Room-Temperature Anomalous Hall Effect, Phys. Rev. Appl. **17**, 064038 (2022).
[35] S. Ding, Y. Peng, M. Xue, Z. Liu, Z. Liang, W. Yang, Y. Sun, J. Zhao, C. Wang, S. Liu, J. Han, and J. Yang, Magnetic phase diagram of $CrPS_4$ and its exchange interaction in contact with NiFe, J. Phys.: Condens. Matter **32**, 405804 (2020).
[36] W. Li, Y. Dai, L. Ni, B. Zhang, D. Tang, Y. Yang, and Y. Xu, Ultrastrong Magnon–Magnon Coupling and Chirality Switching in Antiferromagnet $CrPS_4$, Adv. Funct. Mater. **33**, 2303781 (2023).
[37] J. Son, S. Son, P. Park, M. Kim, Z. Tao, J. Oh, T. Lee, S. Lee, J. Kim, K. Zhang, K. Cho, T. Kamiyama, J. H. Lee, K. F. Mak, J. Shan, M. Kim, J.-G. Park, and J. Lee, Air-Stable and Layer-Dependent Ferromagnetism in Atomically Thin van der Waals $CrPS_4$, ACS Nano **15**, 16904 (2021).
[38] S. Qi *et al.*, Giant electrically tunable magnon transport anisotropy in a van der Waals antiferromagnetic insulator, Nat. Commun. **14**, 2526 (2023).
[39] D. K. de Wal, R. L. Mena, M. Zohaib, and B. J. van Wees, Gate control of magnon spin transport in unconventional magnon transistors based on the van der Waals antiferromagnet $CrPS_4$, Phys. Rev. B **110**, 224434 (2024).
[40] D. K. de Wal, M. Zohaib, and B. J. van Wees, Magnon spin transport in the van der Waals antiferromagnet $CrPS_4$ for noncollinear and collinear magnetization, Phys. Rev. B **110**, 174440 (2024).
[41] X. He, S. Ding, H. G. Giil, J. Wang, M. Bhukta, M. Wu, W. Shi, Z. Lin, Z. Liang, J. Yang, M. Kläui, A. Brataas, Y. Hou, and R. Wu, Spin Seebeck in the weakly exchange-coupled Van der Waals antiferromagnet across the spin-flip transition, Nat. Commun. **16**, 3037 (2025).
[42] See Supplemental Material at [link to be added] for details about sample preparation, magnetization, resistance, field and angular dependence USMR measurement, and model to calculate interfacial spin current $J_s$ between Pt and $CrPS_4$ based on Fermi's golden rule, which includes Refs. [11,31,32,41,43-46,48].
[43] J. Lee, T. Y. Ko, J. H. Kim, H. Bark, B. Kang, S.-G. Jung, T. Park, Z. Lee, S. Ryu, and C. Lee, Structural and Optical Properties of Single- and Few-Layer Magnetic Semiconductor $CrPS_4$, ACS Nano **11**, 10935 (2017).
[44] T. Kikkawa and E. Saitoh, Spin Seebeck Effect: Sensitive Probe for Elementary Excitation, Spin Correlation, Transport, Magnetic Order, and Domains in Solids, Annu. Rev. Condens. Matter Phys. **14**, 129 (2023).
[45] P. Tang and G. E. W. Bauer, Thermal and Coherent Spin Pumping by Noncollinear Antiferromagnets, Phys. Rev. Lett. **133**, 036701 (2024).
[46] C. O. Avci, K. Garello, M. Gabureac, A. Ghosh, A. Fuhrer, S. F. Alvarado, and P. Gambardella, Interplay of spin-orbit torque and thermoelectric effects in ferromagnet/normal-metal bilayers, Phys. Rev. B **90**, 224427 (2014).
[47] S. Ding, P. Noël, G. K. Krishnaswamy, and P. Gambardella, Unidirectional orbital magnetoresistance in light-metal--ferromagnet bilayers, Phys. Rev. Res. **4**, L032041 (2022).
[48] X.-g. Wang, Z.-w. Zhou, Y.-z. Nie, Q.-l. Xia, and G.-h. Guo, Self-consistent study of local and nonlocal magnetoresistance in a YIG/Pt bilayer, Phys. Rev. B **97**, 094401 (2018).

[49] L. Jia, Z. Zheng, X. Zhang, Q. Zhang, S. Sim, T. Zhao, Y. Gu, T. Zeng, R. Xiao, S. Shi, J. Wu, L. Shen, K. Novoselov, and J. Chen, Unidirectional magnetoresistance in the van der Waals antiferromagnet $CrPS_4$, Phys. Rev. B **110**, 214411 (2024).
[50] L. H. Bennett and E. D. Torre, The chemical potential of magnons in quasi-equilibrium, Physica B **403**, 324 (2008).
[51] L. J. Cornelissen, J. Liu, R. A. Duine, J. B. Youssef, and B. J. van Wees, Long-distance transport of magnon spin information in a magnetic insulator at room temperature, Nat. Phys. **11**, 1022 (2015).
[52] L. J. Cornelissen, K. J. H. Peters, G. E. W. Bauer, R. A. Duine, and B. J. van Wees, Magnon spin transport driven by the magnon chemical potential in a magnetic insulator, Phys. Rev. B **94**, 014412 (2016).
[53] S. M. Rezende, R. L. Rodríguez-Suárez, R. O. Cunha, A. R. Rodrigues, F. L. A. Machado, G. A. Fonseca Guerra, J. C. Lopez Ortiz, and A. Azevedo, Magnon spin-current theory for the longitudinal spin-Seebeck effect, Phys. Rev. B **89**, 014416 (2014).
[54] K. S. Olsson, K. An, G. A. Fiete, J. Zhou, L. Shi, and X. Li, Pure Spin Current and Magnon Chemical Potential in a Nonequilibrium Magnetic Insulator, Phys. Rev. X **10**, 021029 (2020).
[55] S. M. Wu, W. Zhang, A. Kc, P. Borisov, J. E. Pearson, J. S. Jiang, D. Lederman, A. Hoffmann, and A. Bhattacharya, Antiferromagnetic Spin Seebeck Effect, Phys. Rev. Lett. **116**, 097204 (2016).
[56] S. M. Rezende, R. L. Rodríguez-Suárez, and A. Azevedo, Theory of the spin Seebeck effect in antiferromagnets, Phys. Rev. B **93**, 014425 (2016).
[57] S. P. Bayrakci, D. A. Tennant, P. Leininger, T. Keller, M. C. R. Gibson, S. D. Wilson, R. J. Birgeneau, and B. Keimer, Lifetimes of Antiferromagnetic Magnons in Two and Three Dimensions: Experiment, Theory, and Numerics, Phys. Rev. Lett. **111**, 017204 (2013).

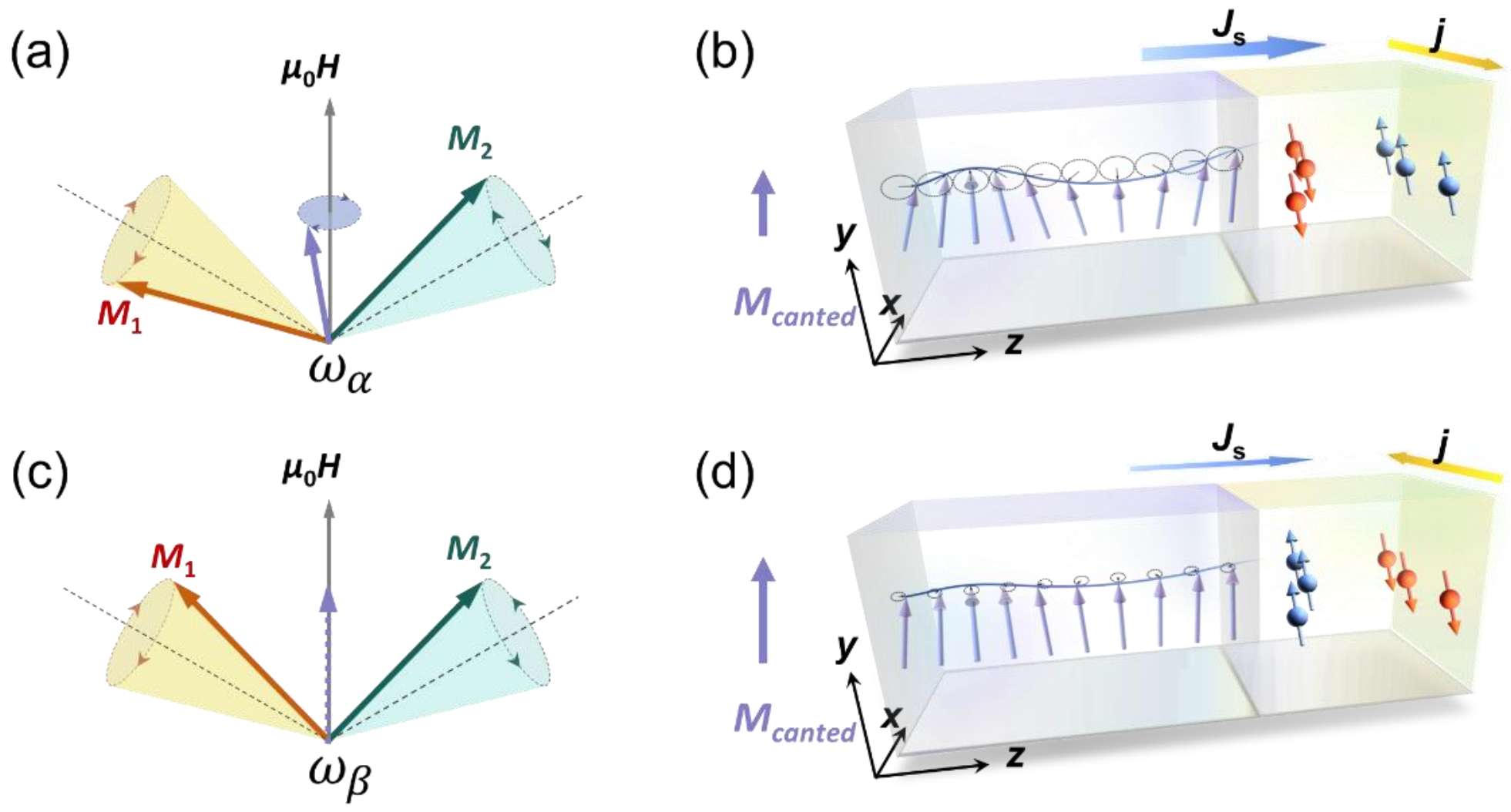

FIG. 1. Spin dynamics and nonequilibrium magnon generation in a canted antiferromagnet. (a) and (c) In the canted state, the net magnetization associated with the $\alpha$ mode precesses around the applied field, while that of the $\beta$ mode oscillates primarily along the field direction. (b) and (d) A nonequilibrium spin accumulation at the interface establishes a finite magnon chemical potential gradient, leading to magnon creation (b) or annihilation (d) depending on the relative orientation between the injected spin polarization and the net magnetization. The magnitude of the canted local magnetization decreases with magnon creation and increases with magnon annihilation. The magnetic moment is opposite in direction to the spin moment. $J_s$ denotes the interfacial spin current, and $j$ is the applied charge current density.

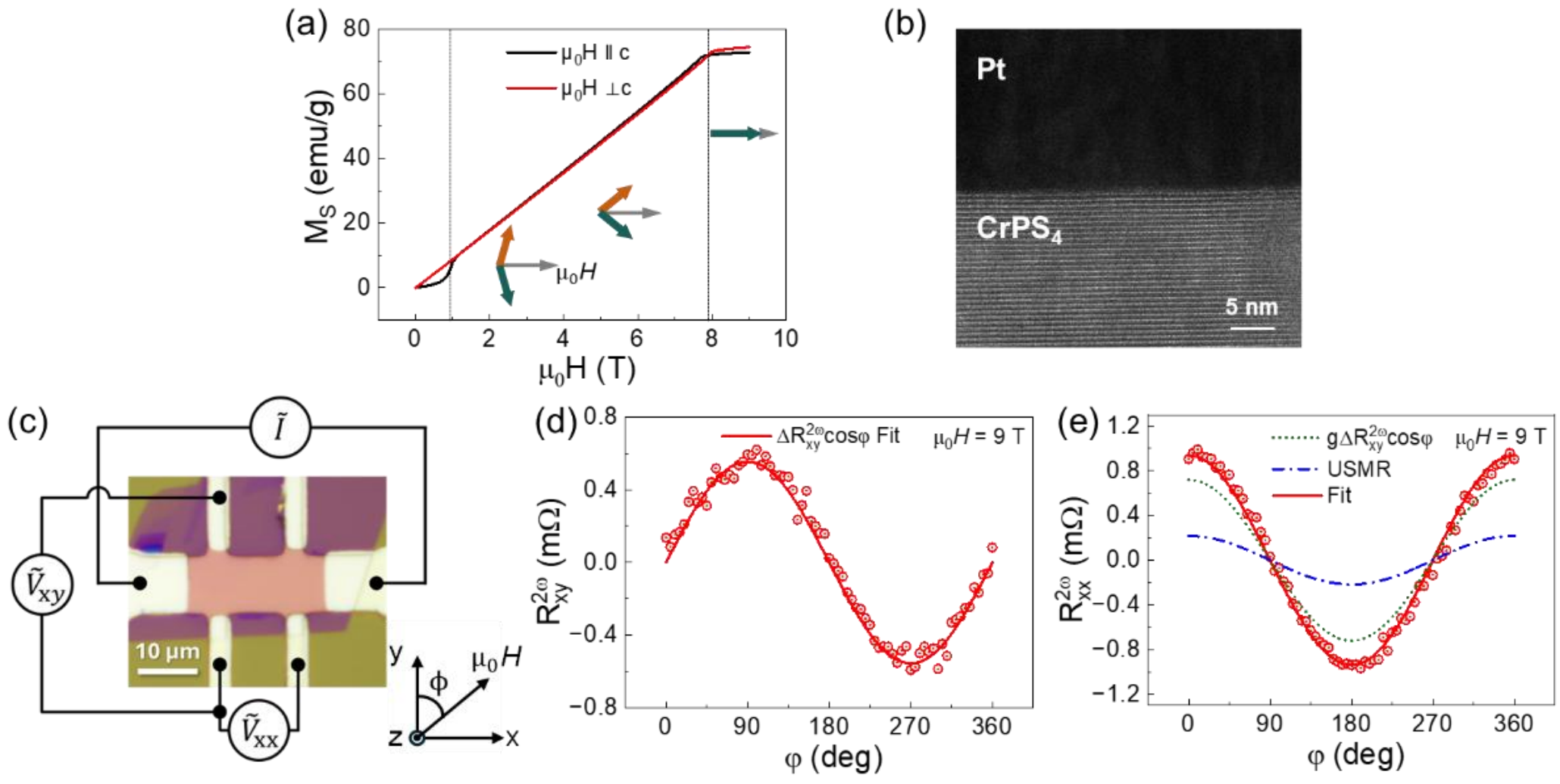

FIG. 2. (a) Magnetic measurements of a $CrPS_4$ single crystal at 10 K show both spin-flop and spin-flip transitions with the magnetic field applied along the *c*-axis, while only the spin-flip transition is observed for fields perpendicular to the *c*-axis. (b) High-resolution cross-sectional TEM image of the $CrPS_4$/Pt interface. (c) Optical micrograph of a $CrPS_4$/Pt Hall bar device on a Si/$SiO_2$ substrate. (d) and (e) Angular dependence of $R_{\mathrm{xy}}^{2\omega}$ and $R_{\mathrm{xx}}^{2\omega}$, respectively, measured in the *xy*-plane under a 9 T field at 5 K with an applied current of 1 mA (peak). The red solid curve in (d) fits a sin$\varphi$ dependence attributed to the spin Seebeck effect. In (e), the blue short dash-dot and green dotted curves represent fitting components from USMR and SSE contribution to $R_{\mathrm{xx}}^{2\omega}$, respectively, and the red solid curve is their sum.

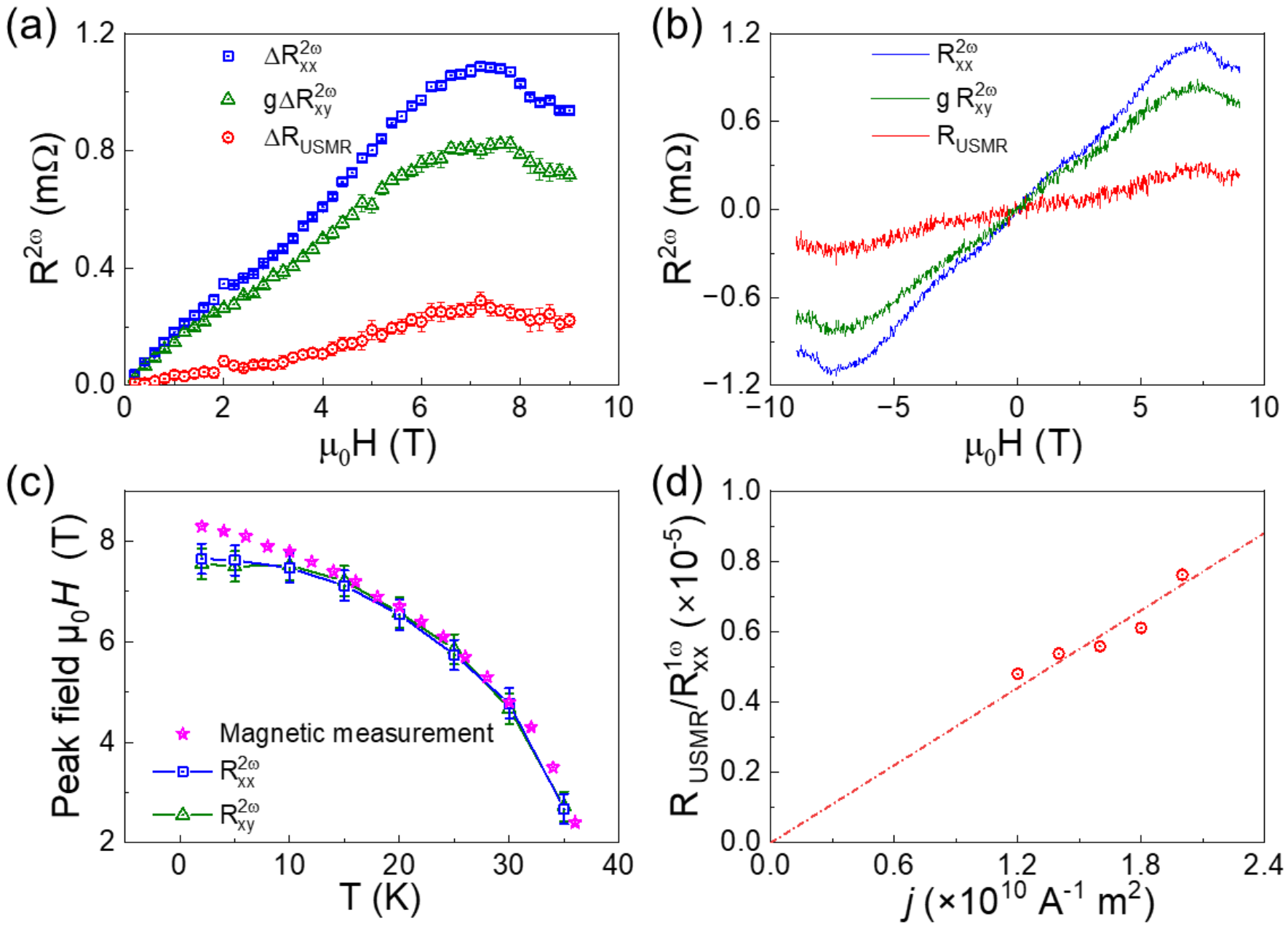

FIG. 3. (a) Field dependence of $\Delta R_{\mathrm{xx}}^{2\omega}$, $g\Delta R_{\mathrm{xy}}^{2\omega}$ and $\Delta R_{USMR} = \Delta R_{\mathrm{xx}}^{2\omega} - g\Delta R_{\mathrm{xy}}^{2\omega}$ at 5 K under an applied current of 1 mA, extracted from fitting of angular dependence measurements performed at various magnetic fields [42]. $g$ is the geometric factor. (b) Field dependence of $R_{\mathrm{xx}}^{2\omega}$ and $gR_{\mathrm{xy}}^{2\omega}$ obtained from measurements with the magnetic field applied along the *y*-axis and *x*-axis, and $R_{USMR} = R_{\mathrm{xx}}^{2\omega} - gR_{\mathrm{xy}}^{2\omega}$. (c) $R_{\mathrm{xx}}^{2\omega}$ and $R_{\mathrm{xy}}^{2\omega}$ peak field as a function of temperature (blue square and triangle green), which is similar to the temperature dependence of the spin-flip transition field (magenta star). (d) Variation of the $\Delta R_{USMR}/R_{\mathrm{xx}}^{1\omega}$ as a function of applied current, obtained at 5 K under an 8 T magnetic field. Here, $\Delta R_{USMR} = (R_{USMR}(8\,T) - R_{USMR}(-8\,T))/2$ from the field dependent measurement. The solid line is a linear fit to the data constrained through the origin.

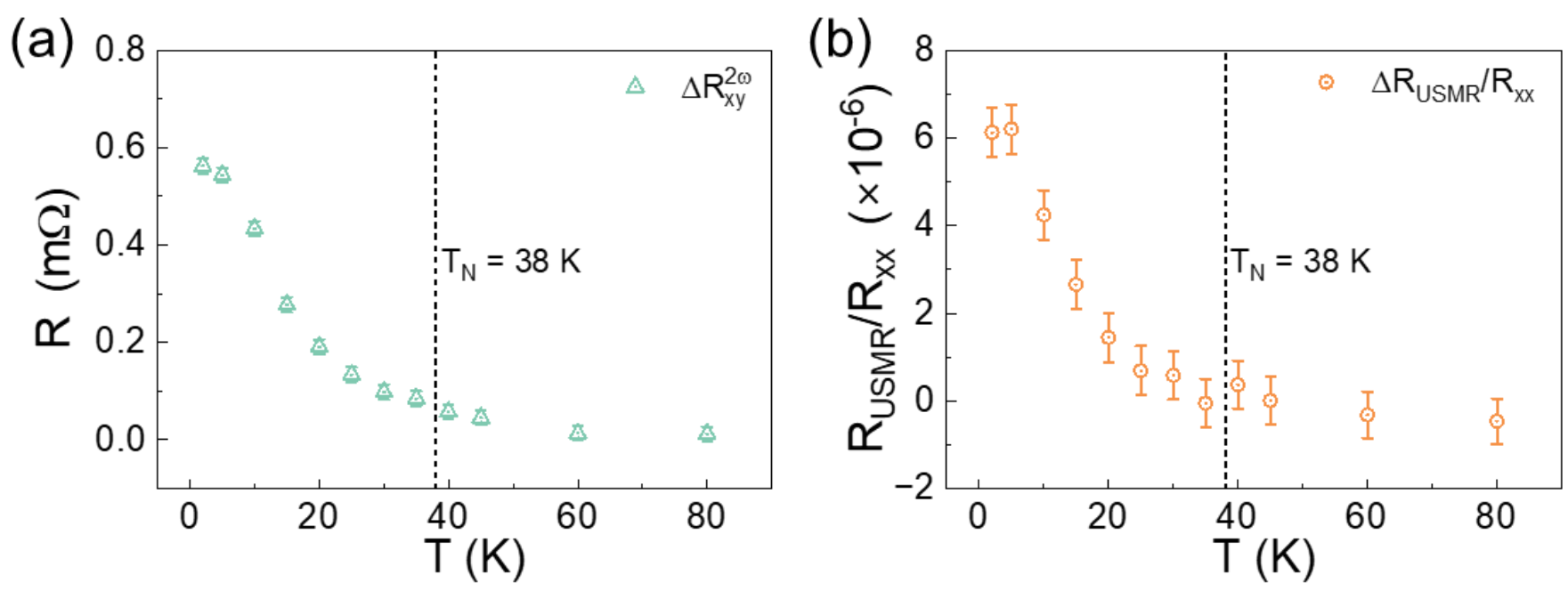

FIG. 4. (a) Temperature dependence of the transverse spin Seebeck resistance $\Delta R_{\mathrm{xy}}^{2\omega}$, extracted from the second-harmonic transverse voltage. (b) USMR magnitude $\Delta R_{USMR}/R_{xx}$ plotted as functions of temperature.